\documentclass[sigconf]{acmart}

\newcommand{\delete}[1]{}%

\usepackage{enumitem}

\usepackage{xspace} %

\usepackage{hyperref}
\usepackage[capitalise]{cleveref}
\crefname{table}{Tab.}{Tabs.}
\Crefname{table}{Table}{Tables}
\crefname{section}{Sec.}{Secs.}
\Crefname{section}{Section}{Sections}
\Crefname{figure}{Fig.}{Fig.}
\Crefname{figure}{Figure}{Figures}

\newcommand{\quotedfeedback}[1]{{\textit{#1}}}

\newcommand{\lfSpecSymbol}[1]{\texttt{#1}}
\newcommand{\lfSpecType}[1]{\textit{#1}}
\newcommand{\lfSpecTarget}[1]{\textit{#1}}
\newcommand{\lfSpecCategory}[1]{\textit{#1}}
\newcommand{\lfSpecAggregationMethod}[1]{\textit{#1}}

\newcommand{\evaluatedMethod}[1]{\textit{#1}}
\newcommand{\evaluatedMethodDoccano}{\evaluatedMethod{Doccano}\xspace}
\newcommand{\evaluatedMethodSnorkel}{\evaluatedMethod{Snorkel}\xspace}
\newcommand{\evaluatedMethodDALL}{\evaluatedMethod{DALL}\xspace}
\newcommand{\evaluatedMethodReusedDALL}{\evaluatedMethod{Reused DALL}\xspace}
\newcommand{\evaluatedMethodGPT}{\evaluatedMethod{GPT}\xspace}
\newcommand{\evaluatedMethodDP}{\evaluatedMethod{DP}\xspace}
\newcommand{\evaluatedMethodDPAL}{\evaluatedMethod{DP+AL}\xspace}
\newcommand{\evaluatedMethodDPALLLM}{\evaluatedMethod{DP+AL+LLM}\xspace}

\AtBeginDocument{%
  }

\copyrightyear{2026}
\acmYear{2026}
\setcopyright{cc}
\setcctype{by}
\acmConference[CHI '26]{Proceedings of the 2026 CHI Conference on Human Factors in Computing Systems}{April 13--17, 2026}{Barcelona, Spain}
\acmBooktitle{Proceedings of the 2026 CHI Conference on Human Factors in Computing Systems (CHI '26), April 13--17, 2026, Barcelona, Spain}
\acmDOI{10.1145/3772318.3791356}
\acmISBN{979-8-4007-2278-3/2026/04}

\begin{document}

\title{DALL: Data Labeling via Data Programming and Active Learning Enhanced by Large Language Models}

\author{Guozheng Li}
\orcid{0000-0001-6663-6712}
\affiliation{%
  \institution{Beijing Institute of Technology}
  \city{Beijing}
  \country{China}
}
\email{guozheng.li@bit.edu.cn}

\author{Ao Wang}
\orcid{0009-0008-8569-4473}
\affiliation{%
  \institution{Beijing Institute of Technology}
  \city{Beijing}
  \country{China}
}
\email{3220241402@bit.edu.cn}

\author{Shaoxiang Wang}
\orcid{0009-0003-6274-2958}
\affiliation{%
  \institution{Beijing Institute of Technology}
  \city{Beijing}
  \country{China}
}
\email{3220241408@bit.edu.cn}

\author{Yu Zhang}
\orcid{0000-0002-9035-0463}
\authornote{Yu Zhang is the corresponding author}
\affiliation{%
  \institution{University of Oxford}
  \city{Oxford}
  \country{United Kingdom}
}
\email{yuzhang94@outlook.com}

\author{Pengcheng Cao}
\orcid{0009-0006-2007-6944}
\affiliation{%
  \institution{People's Daily}
  \city{Beijing}
  \country{China}
}
\email{pengchengcao24@163.com}

\author{Yang Bai}
\orcid{0009-0006-7589-6114}
\affiliation{%
  \institution{People's Daily}
  \city{Beijing}
  \country{China}
}
\email{yangbai_2024@163.com}

\author{Chi Harold Liu}
\orcid{0000-0002-0252-329X}
\affiliation{%
  \institution{Beijing Institute of Technology}
  \city{Beijing}
  \country{China}
}
\email{liuchi02@gmail.com}

\begin{abstract}
  Deep learning models for natural language processing rely heavily on high-quality labeled datasets.
However, existing labeling approaches often struggle to balance label quality with labeling cost.
To address this challenge, we propose DALL, a text labeling framework that integrates data programming, active learning, and large language models.
DALL introduces a structured specification that allows users and large language models to define labeling functions via configuration, rather than code.
Active learning identifies informative instances for review, and the large language model analyzes these instances to help users correct labels and to refine or suggest labeling functions.
We implement DALL as an interactive labeling system for text labeling tasks.
Comparative, ablation, and usability studies demonstrate DALL's efficiency, the effectiveness of its modules, and its usability.

\end{abstract}




\maketitle

\section{Introduction}
\label{sec:introduction}

Deep learning models have achieved remarkable success in natural language processing (NLP) tasks such as document classification and sentiment analysis, but their effectiveness depends on large, high-quality labeled datasets.
Constructing such datasets often requires substantial human effort~\cite{Chang2017Revolt}, making data labeling a critical bottleneck for the continued advancement of deep learning~\cite{Sun2017Revisiting}.
Manual labeling by human annotators can produce high-quality labels but is labor-intensive, time-consuming, and costly~\cite{Nashaat2018Hybridization, Chegini2019Interactive}.

To address these challenges, various techniques have been introduced to accelerate data labeling and reduce manual effort.
Data programming (DP)~\cite{Ratner2016Data} lets users encode domain knowledge as labeling functions, which can then be applied to assign labels to large datasets.
However, it demands programming skill and is often limited by accuracy.
Active learning (AL)~\cite{Settles2009Active} lowers labeling cost by selecting the most informative samples for annotation, improving model performance with fewer labels.
It nevertheless suffers from a cold-start problem when initial labels are insufficient.
More recently, large language models (LLMs) have demonstrated promising capabilities to act as annotators in labeling tasks~\cite{Gilardi2023ChatGPT}, yet small models may outperform them with limited labeled data~\cite{Lu2023Human}, and using LLMs may involve trial-and-error for prompt engineering.

To address these limitations and achieve a better balance between labeling efficiency and accuracy, we propose DALL, a data labeling framework that combines \underline{D}ata programming and \underline{A}ctive learning, enhanced by \underline{L}arge \underline{L}anguage models.
The three components complement each other.
Data programming supplies large-scale, albeit noisy, labels and thus alleviates the cold start of active learning.
Active learning directs human and model effort to the most informative instances, allowing fine-grained refinement of data labels.
LLMs assist experts in correcting labels and refining labeling functions, reducing the burden of manual revision.
Through an iterative process that utilizes the three components, DALL helps refine data labels efficiently and progressively.
We implemented the DALL labeling system based on the DALL framework to validate its effectiveness.

To support efficient and principled authoring of labeling functions, DALL introduces a structured specification that enables both users and LLMs to define labeling rules through configuration rather than manual coding.
This specification provides an interpretable representation of labeling strategies, allowing users to clearly express labeling targets and rules.
It enables labeling rules to be reused and continuously improved across tasks and provides traceability.

We conducted a comparative study to evaluate the effectiveness of the labeling system against widely used alternatives: Doccano~\cite{Nakayama2018doccano}, Snorkel~\cite{Ratner2017Snorkel}, and LLM-based labeling with GPT-3.5 Turbo~\cite{OpenAI2023GPT35}.
The results show that our system achieved comparable or higher accuracy with substantially lower labeling time.
We also conducted an ablation study to evaluate the contributions of DALL's three major modules: data programming, active learning, and LLM.
The results demonstrate that the modules complement each other effectively, jointly enhancing labeling accuracy.
We collected usability feedback from participants after the comparative and ablation studies.
The feedback confirmed the system's usability and practical value.

This work has the following contributions:

\begin{itemize}[leftmargin=3.5mm]
    \item We propose DALL, a data labeling framework that combines data programming, active learning, and large language models to achieve a better trade-off between efficiency and label quality.

    \item We implement the DALL labeling system for text labeling tasks based on the DALL framework.
          Through comparative, ablation, and usability studies, we demonstrate the effectiveness of the labeling system.
\end{itemize}

DALL's source code is available at \url{https://github.com/bitvis2021/DALL}.

\section{Related Work}

DALL builds on data programming, active learning, and LLMs, and unifies them in a labeling framework.
We review prior work in each of these areas to position our design choices, then discuss frameworks that combine them, which motivates our integration of all three.

\subsection{Data Programming}
\label{sec:data-programming}

Conventionally, data labeling is performed instance by instance, where annotators label data objects individually or in small batches.
This approach is referred to as \textit{instance-based labeling}, and is common in systems such as Doccano~\cite{Nakayama2018doccano} and Label Studio~\cite{Tkachenko2020Label}.
Instance-based labeling works well with techniques such as default labels and active learning~\cite{Zhang2024Simulation}.

In contrast, data programming enables users to encode domain knowledge as programmatic labeling functions, which a generative model aggregates to produce large-scale, potentially noisy labels~\cite{Ratner2016Data}.
It scales efficiently but does not offer the same per-instance control as instance-based labeling.
Snorkel~\cite{Ratner2017Snorkel} is a representative system that treats labeling functions as weak supervision to train such a model to label entire datasets.
Despite the effectiveness of data programming, its usage is often hindered by the programming expertise required to implement labeling functions.

To reduce the programming burden, Ruler~\cite{Evensen2020Ruler} introduces data programming by demonstration (DPBD), synthesizing labeling functions from span-level interactive demonstrations on document examples for document-level classification.
TagRuler~\cite{Choi2021TagRuler} further extends DPBD to span-level labeling tasks.
Ruler and TagRuler reduce the programming burden while achieving model quality comparable to manual data programming.

Further work combines visualization and DPBD in data programming for other data modalities.
Hoque et al.~\cite{Hoque2023Visual} introduce visual concept programming for semantic segmentation of images at scale.
He et al.~\cite{He2024VideoPro} propose VideoPro for video, using events as building blocks for labeling functions and supporting iterative refinement of templates.
Li et al.~\cite{Li2025EvoVis} present EvoVis to understand labeling iterations.
He et al.~\cite{He2025ProTAL} propose ProTAL, a drag-and-link framework for temporal action localization where users define actions by linking body parts and objects with spatial relationships.

DALL adopts the data-programming paradigm for scale but addresses the programming burden through a structured specification that allows users and LLMs to define labeling functions via configuration rather than code.
It couples data programming with active learning to refine noisy labels.

\subsection{Active Learning}
\label{sec:active-learning}

Active learning~\cite{Settles2009Active} improves labeling efficiency by selecting the most informative instances for human annotation.
Selection strategies typically measure informativeness via criteria such as disagreement among a committee of models~\cite{Seung1992Query}, label uncertainty~\cite{Lewis1994Heterogeneous}, and data density~\cite{Nguyen2004Active}.
Some labeling systems combine these strategies with interactive visualizations, allowing algorithms and annotators to jointly choose which instances to label~\cite{Liao2016Visualization}.

Unlike conventional active learning settings that query new labels for unlabeled data, another line of work employs active learning to refine noisy labels.
Bernhardt et al.~\cite{Bernhardt2022Active} rank instances by estimated label error and re-labeling difficulty to prioritize which samples to re-annotate.
Yuan et al.~\cite{Yuan2024Hide} use LLMs as active annotators to correct noisy labels, partitioning data by estimated noise level and querying the LLM for label corrections on noisy subsets.

Similarly, DALL uses active learning to refine the labels produced by the labeling functions.
It follows the uncertainty-based sampling framework~\cite{Lewis1994Heterogeneous} to prioritize instances with the least model confidence and the greatest labeling function disagreement for review and refinement, as described in \cref{sec:active-learning-module}.

\subsection{LLM for Data Labeling}

LLMs have demonstrated promising capabilities in many tasks~\cite{Aodeng2025InReAcTable, Wang2024Visualization}.
Their progress has led to increasing use as annotators for data labeling.
The human's role increasingly focuses on validating or correcting LLM-generated labels rather than generating all labels from scratch~\cite{Zhang2025Position}.

Nevertheless, fully automated LLM-based labeling often produces unreliable or inconsistent annotations compared to human labeling, underscoring the need for human oversight~\cite{Mohta2023Are}.
Empirical studies~\cite{Lu2023Human} have demonstrated that, although LLMs such as GPT-3.5 can serve labeling purposes, active learning with small models rapidly outperforms LLMs in annotation quality after only a small amount of expert-labeled data.
These findings highlight both the effectiveness of using active learning to iteratively improve smaller models and the indispensable role of expert annotators.
Furthermore, over-reliance on LLM-generated labels can introduce automation bias~\cite{Goddard2012Automation}, which can potentially lead to reduced annotation quality.
These limitations suggest the need for human oversight.
In DALL, we thus use LLMs to assist in correcting labels and refining labeling functions, rather than as the sole annotator.

LLMs can also serve roles beyond direct annotation (e.g., assisting heuristic design or verification), as will be explored in \cref{sec:data-labeling-frameworks}.

\subsection{Data Labeling Frameworks}
\label{sec:data-labeling-frameworks}

Prior work has developed labeling frameworks that combine different techniques (e.g., multiple instance learning with active learning~\cite{Wang2011Active}, active learning with semi-supervised learning~\cite{Zhang2021MI3}) and explored the design space of labeling systems~\cite{Zhang2022OneLabeler}.
Of direct relevance to DALL, prior work has explored pairwise combinations of data programming, active learning, and LLM-based labeling.

\textbf{Data Programming + Active Learning.}
Existing approaches that combine data programming and active learning either use data programming to provide initial labels for active learning, or use active learning for refining labeling functions in data programming.
Nashaat et al.~\cite{Nashaat2018Hybridization} employ data programming to generate labeling functions to produce preliminary coarse labels, thereby alleviating the cold start problem in active learning.
They then use active learning to refine these noisy labels, achieving higher quality while reducing manual labeling costs.
In contrast, Boecking et al.~\cite{Boecking2021Interactive} introduce IWS, where the system proposes labeling functions, the user gives feedback, and the system uses feedback to learn which are effective and to guide further proposals.
Hsieh et al.~\cite{Hsieh2022Nemo} present Nemo, which uses active learning to select development data for deriving labeling functions and contextualizes labeling functions by modeling each LF's creation context to decide where it applies.
Guan et al.~\cite{Guan2024ActiveDP} use active learning to select instances for developing labeling functions, then apply data programming at scale.
DALL follows the line of using active learning to refine labels generated by labeling functions and additionally integrates LLM assistance for authoring and refining labeling functions.

\textbf{Data Programming + LLMs.}
Recent work uses LLMs to generate or act as labeling functions in data programming, reducing the need for hand-written rules.
Li et al.~\cite{Li2024LLMassisted} apply programmatic weak supervision to semantic type detection in data lake tables.
They use LLMs to generate labeling functions and explore prompt engineering strategies to address the difficulty of manually writing functions.
Su et al.~\cite{Su2023Leveraging} propose prompted weak supervision, using LLMs as labeling functions and introducing a structure refining module that learns dependencies among prompted sources from prompt similarities in embedding space.

\textbf{Active Learning + LLMs.}
Recent work has combined active learning with LLMs.
For example, MEGAnno+~\cite{Kim2024MEGAnno} uses LLMs to create labels that annotators selectively verify according to active learning strategies.
Wang et al.~\cite{Wang2024HumanLLM} propose a workflow where LLMs generate labels and explanations, a verifier assesses label quality, and active learning directs human effort to uncertain cases.

We observe that existing labeling approaches have exploited complementarities between pairs of data programming, active learning, and LLMs, while overlooking the potential synergy of combining all three.
In this work, we propose DALL, a unified framework that integrates all three to enhance labeling efficiency and accuracy.

\section{Preliminary Study}
\label{sec:preliminary-study}

We settled on improving data programming and conducted a preliminary study to better understand requirements for text labeling systems and challenges in data programming.
This section describes the study design and the design requirements we derived from participants' feedback.

\subsection{Study Design}
We describe the participants, procedure, and feedback.

\textbf{Participants.}
The preliminary study involved five participants: two professors with research experience in data labeling (E1 and E2), two fourth-year graduate students (E3 and E4), and a software engineer with three years of engineering experience (E5).

\textbf{Procedure.}
Following the categorization in \cref{sec:data-programming}, labeling systems typically adopt either instance-based labeling or data programming.
We selected four open-source systems spanning both paradigms: Doccano~\cite{Nakayama2018doccano} and Label Studio~\cite{Tkachenko2020Label} for instance-based labeling, and Snorkel~\cite{Ratner2017Snorkel} and Ruler~\cite{Evensen2020Ruler} for data programming.
Participants were introduced to these systems through walk-through demonstrations of their labeling workflows and engaged in semi-structured interviews, reflecting on their experiences and identifying the most time-consuming and labor-intensive parts of the labeling process.

\textbf{Feedback.}
The following presents excerpts of the key feedback from the demonstrations.
E5 suggested that ``\quotedfeedback{although the individual labeling process is time-consuming for large datasets, it is still essential to provide instance-based labeling in the labeling system, for users to directly edit and validate the labeling results and ensure label quality}'' \textit{(comment 1)}.
E1 highlighted the efficiency of data programming, noting that ``\quotedfeedback{data programming can significantly save time}''.
However, he also emphasized that ``\quotedfeedback{the labeling process requires repeatedly inspecting samples and writing code, which increases the programming workload}'' \textit{(comment 2)}.
Similarly, E2 pointed out that ``\quotedfeedback{Snorkel lacks an effective way to create labeling functions, limiting the ability to efficiently evaluate and refine them}'' \textit{(comment 3)}.
Participants commented that Ruler's demonstration-based interface relieved the programming burden compared to Snorkel and provided a more intuitive view to inspire heuristics.
However, E3 pointed out that ``\quotedfeedback{Ruler's interface only displays individual data objects along with a statistical overview of labeling functions, making it difficult to gain an intuitive understanding of their overall effects}'' \textit{(comment 4)}.
E4 further noted that ``\quotedfeedback{within this interface, experts must sequentially switch between examples to construct labeling functions, which is time-consuming when creating numerous labeling functions and limits the flexibility to iteratively refine them}'' \textit{(comment 5)}.

\begin{figure*}
  \centering
  \includegraphics[width=\linewidth]{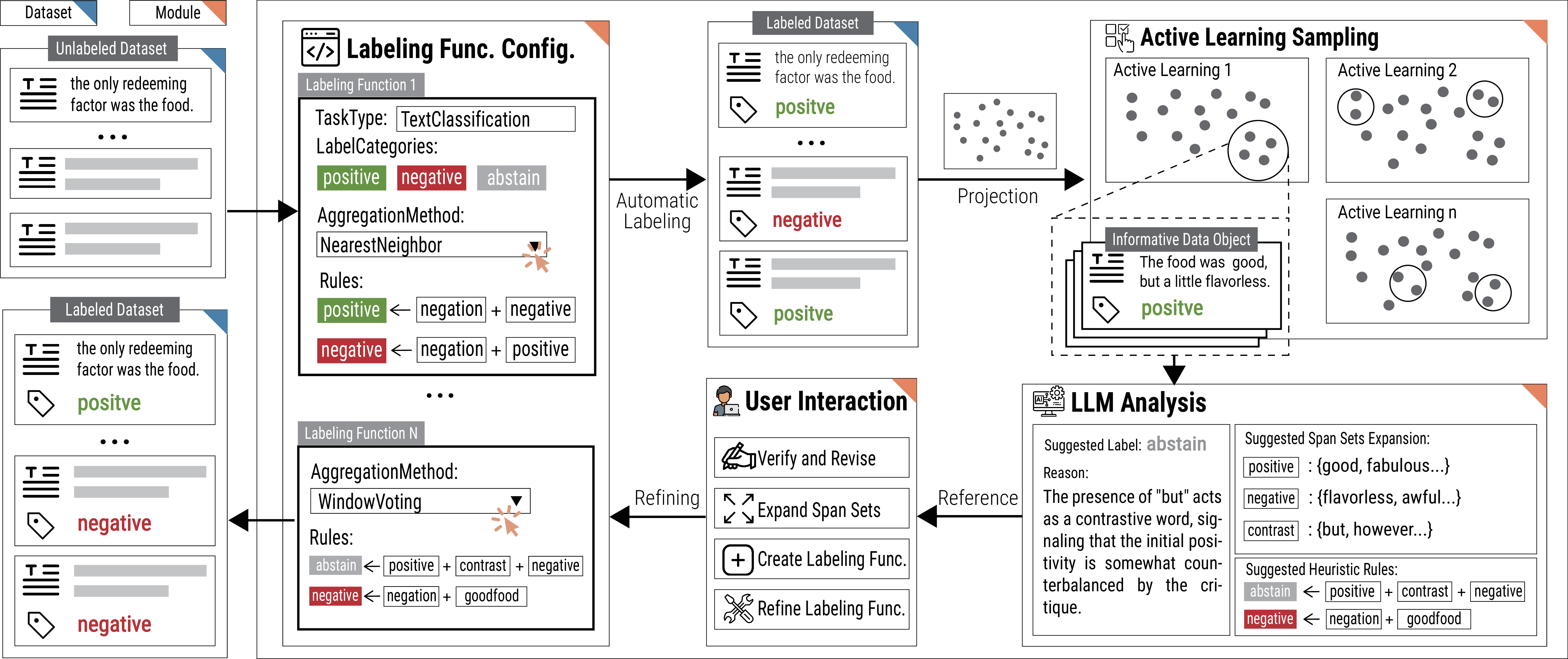}
  \caption{
    The DALL Framework.
    Given an unlabeled dataset, users can create labeling functions to annotate it.
    DALL uses active learning algorithms to select informative samples and provides LLM-powered analysis to assist users in refining their labeling functions.
    Its goal is to iteratively enhance labeling quality.
  }
  \Description{
    A workflow diagram of the DALL Framework.
    It shows a central dataset represented by a blue triangle, connected in a loop to three orange-colored modules: labeling function creation, active learning algorithm selection, and LLM-powered analysis.
    Arrows indicate an iterative process where data is processed to enhance labeling quality.
  }
  \label{fig:framework}
\end{figure*}

\subsection{Design Requirements}

Based on participant feedback and our review of prior research, we distilled the following three requirements, which informed the design of the DALL framework.

\begin{itemize}[leftmargin=6mm]
  \item [\textbf{R1}] \textbf{Lower the barrier for authoring labeling functions.}
        While data programming can be efficient, it typically requires repeatedly inspecting samples and writing code, which can be a significant barrier for many annotators (\textit{comment 2}).
        E2 pointed out that current data programming systems lack an effective way to create and refine labeling functions (\textit{comment 3}).
        In line with this feedback, the framework should offer an interactive, no-code interface that enables users to articulate domain knowledge and heuristics and to convert them into labeling functions without programming.

  \item [\textbf{R2}] \textbf{Combine data programming and instance-based labeling.}
        Data programming can rapidly scale label generation but often produces inaccurate results.
        E5 emphasized that instance-based labeling remains essential for users to directly validate and edit individual labels when necessary (\textit{comment 1}).
        Therefore, the framework should couple data programming with instance-level feedback to allow fine-grained control.

  \item [\textbf{R3}] \textbf{Enable efficient iteration with guided refinement.}
        Ambiguities in text frequently lead to noisy initial labels, and iterative correction substantially improves label quality.
        However, users may struggle to prioritize which items to review.
        E3 (\textit{comment 4}) and E4 (\textit{comment 5}) highlighted that existing labeling systems make it difficult to gain an intuitive understanding of labeling functions' overall effects and to iteratively refine them.
        Therefore, the framework should provide guided refinement by surfacing high-priority samples or common labeling errors so users can focus on the most impactful corrections.

\end{itemize}

\section{DALL Framework}
\label{sec:framework}

DALL focuses on two types of text labeling tasks, text classification and target-specific classification, as formalized below.

\begin{itemize}[leftmargin=3.5mm]
  \item \textbf{Text classification} is to assign a single label to each text.
  \begin{itemize}[leftmargin=*]
    \item \textbf{Input}: a corpus $X = (x_1, \dots, x_n)$ where each $x_i$ is a string.
    \item \textbf{Output}: for each $x_i$, a label $y_i \in \mathcal{Y}$ where $\mathcal{Y}$ is the set of label categories.
  \end{itemize}
  
  \item \textbf{Target-specific classification} is to assign a single label to each target in a text. Aspect-based sentiment analysis is an example and widely used in news analysis~\cite{BiaSeer2025}. 
  \begin{itemize}[leftmargin=*]
    \item \textbf{Input}: a corpus $X = (x_1, \dots, x_n)$ and a set of targets $T = (t_1, \dots, t_m)$ (each $t_j$ is a string). For each text $x_i$, only a subset $T_i \subseteq T$ of targets may be discussed.
    \item \textbf{Output}: for each $x_i$, a label $y_{ij} \in \mathcal{Y}$ per target $t_j \in T_i$ where $y_{ij}$ is the label for target $t_j$ in text $x_i$.
  \end{itemize}
\end{itemize}

Note that for both tasks, we focus on single-label classification and do not consider the case where an entity can be associated with multiple labels.

The DALL framework combines data programming, active learning, and LLMs.
Given an unlabeled dataset, DALL enables users to perform data labeling through the loop of three steps (see \cref{fig:framework}).
First, following the paradigm of data programming, users encode their labeling heuristics into labeling functions according to a specification and label the dataset with these functions (\cref{sec:data-programming-module}).
Second, DALL utilizes active learning strategies to identify and select informative instances from the labeled dataset (\cref{sec:active-learning-module}).
Third, LLMs analyze the selected instances to assist in creating effective labeling functions (\cref{sec:llm-module}).
DALL is an iterative framework that allows users to progressively refine the labeling result.

\subsection{Data Programming Module}
\label{sec:data-programming-module}

\begin{figure}
  \centering
  \includegraphics[width=\linewidth]{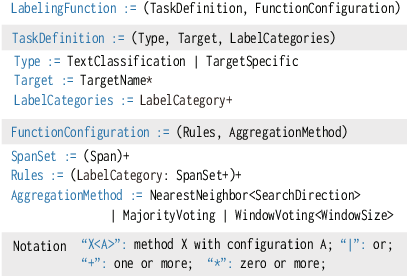}
  \caption{
    The labeling function specification.
    A labeling function consists of two components: task definition and function configuration.
    The task definition specifies the task type and the set of label categories used throughout the labeling process.
    The function configuration assigns specific logic to each function that guides how labels are generated.
  }
  \Description{
    A diagram showing the labeling function specification, divided into two main parts: task definition and function configuration.
  }
  \label{fig:specification}
\end{figure}

Data programming conventionally requires users to implement labeling functions in code, posing a high entry barrier for users without programming expertise.
To address this issue, we propose a structured yet flexible specification that allows users to express heuristics as labeling functions without coding \textbf{(R1)}.
The following focuses on describing the specification.
As \cref{fig:specification} shows, the specification defines a labeling function with two components: task definition and function configuration.

\begin{itemize}[leftmargin=3.5mm]
  \item \textbf{Task definition} specifies the labeling task and label categories (see \cref{sec:task-definition}).
  \item \textbf{Function configuration} describes how labels are generated (see \cref{sec:function-configuration}).
\end{itemize}

\subsubsection{Task Definition}
\label{sec:task-definition}

The task definition includes three attributes: \lfSpecSymbol{Type}, \lfSpecSymbol{Target}, and \lfSpecSymbol{LabelCategories}.

\textbf{\lfSpecSymbol{Type}} specifies the granularity of the labeling task, with two options supported in DALL: \lfSpecType{TextClassification} and \lfSpecType{TargetSpecific}.

\begin{itemize}[leftmargin=3.5mm]
  \item \lfSpecType{TextClassification} corresponds to a coarse-grained task of assigning one label to the entire text (e.g., topic classification).

  \item \lfSpecType{TargetSpecific} corresponds to a fine-grained task of assigning a label to a specified target within the text, such as an entity or aspect (e.g., aspect-based sentiment analysis~\cite{Zhang2023Survey}).
\end{itemize}

\textbf{\lfSpecSymbol{Target}} specifies the data elements to which labels are applied for the \lfSpecType{TargetSpecific} labeling task.
The specified target entities are in string list format so that the labeling functions can recognize and process them within the text.
This attribute is invalid when the \lfSpecSymbol{Type} is \lfSpecType{TextClassification}, in which case \lfSpecSymbol{Target} is regarded as an empty list.
(Type: potentially empty list of \lfSpecSymbol{TargetName}, where each \lfSpecSymbol{TargetName} is a string)

\textbf{\lfSpecSymbol{LabelCategories}} defines the set of possible labels for the labeling task.
(Type: non-empty list of \lfSpecSymbol{LabelCategory}, where each \lfSpecSymbol{LabelCategory} is a string)

\subsubsection{Function Configuration}
\label{sec:function-configuration}

The function configuration determines the labeling functionality of a labeling function and consists of three components: \lfSpecSymbol{SpanSet}, \lfSpecSymbol{Rules}, and \lfSpecSymbol{AggregationMethod}.

\textbf{\lfSpecSymbol{SpanSet}} refers to a collection of semantically related spans.
Each span is one or more consecutive words in a text that together represent a meaningful concept, such as a domain-specific term.
The labeling process begins by grouping relevant spans into span sets.
As will be discussed in \cref{sec:llm-module}, the span sets are created collaboratively by users and LLMs.
The extraction of span sets allows users to define rules based on high-level concepts associated with the span sets.

\textbf{\lfSpecSymbol{Rules}} is a collection of heuristic rules for mapping span sets to label categories.
Each rule maps one or more span sets to a corresponding label category.
For example, the rule $\{Favor: \textit{support}\}$ assigns the label category \lfSpecCategory{Favor} to the span set \textit{support}.
Using the rules, a part of a text (i.e., a \lfSpecSymbol{Span}) can be assigned a label category according to the characteristics of the spans it contains.
In complex scenarios where individual spans provide insufficient context, joint mapping can be used to transform sequential spans into a single label category.
For example, the rule $\{Against:(negation, support)\}$ assigns \lfSpecCategory{Against} when two adjacent spans match the span sets \textit{negation} and \textit{support} in that order.
We do not allow a span to be assigned multiple label categories.
During matching, longer rules are given higher priority, and when multiple rules are of the same length, the rule created later is given higher priority.

\textbf{\lfSpecSymbol{AggregationMethod}} is used to aggregate label categories converted by rules within a text into a single label category for the entire text (i.e., for \lfSpecType{TextClassification} task) or a specified target (i.e., for \lfSpecType{TargetSpecific} task).
A text contains multiple spans, with each span associated with one or more label categories according to \lfSpecSymbol{Rules}.

\begin{itemize}[leftmargin=3.5mm]
  \item \textbf{For \lfSpecType{TextClassification} task.}
    The \lfSpecAggregationMethod{MajorityVoting} aggregation method can be used to aggregate the label categories of spans within a text into a single label category for the entire text by selecting the majority class.
    When multiple label categories have the same highest count, the abstain category is assigned.
  
  \item \textbf{For \lfSpecType{TargetSpecific} task.}
    Two aggregation methods are supported: \lfSpecAggregationMethod{NearestNeighbor} and \lfSpecAggregationMethod{WindowAnalysis}.
    Both methods concentrate on spans in close proximity to the \lfSpecSymbol{Target}.

    \begin{itemize}
      \item \lfSpecAggregationMethod{NearestNeighbor} first identifies the position of the \lfSpecSymbol{Target}, then searches for the closest \lfSpecSymbol{Span} and assigns the span's \lfSpecSymbol{LabelCategory} to the target.
      The search can be configured to accommodate different linguistic structures through the \lfSpecSymbol{SearchDirection} parameter.
      
      \item \lfSpecAggregationMethod{WindowAnalysis} performs majority voting within a local window around the \lfSpecSymbol{Target}, similar to \lfSpecAggregationMethod{MajorityVoting}.
      The window size is configured via \lfSpecSymbol{WindowSize}: setting it to $n$ uses the $n$ spans immediately before and the $n$ spans immediately after the \lfSpecSymbol{Target} for voting.
    \end{itemize}

\end{itemize}

\begin{figure}
  \includegraphics[width=\columnwidth]{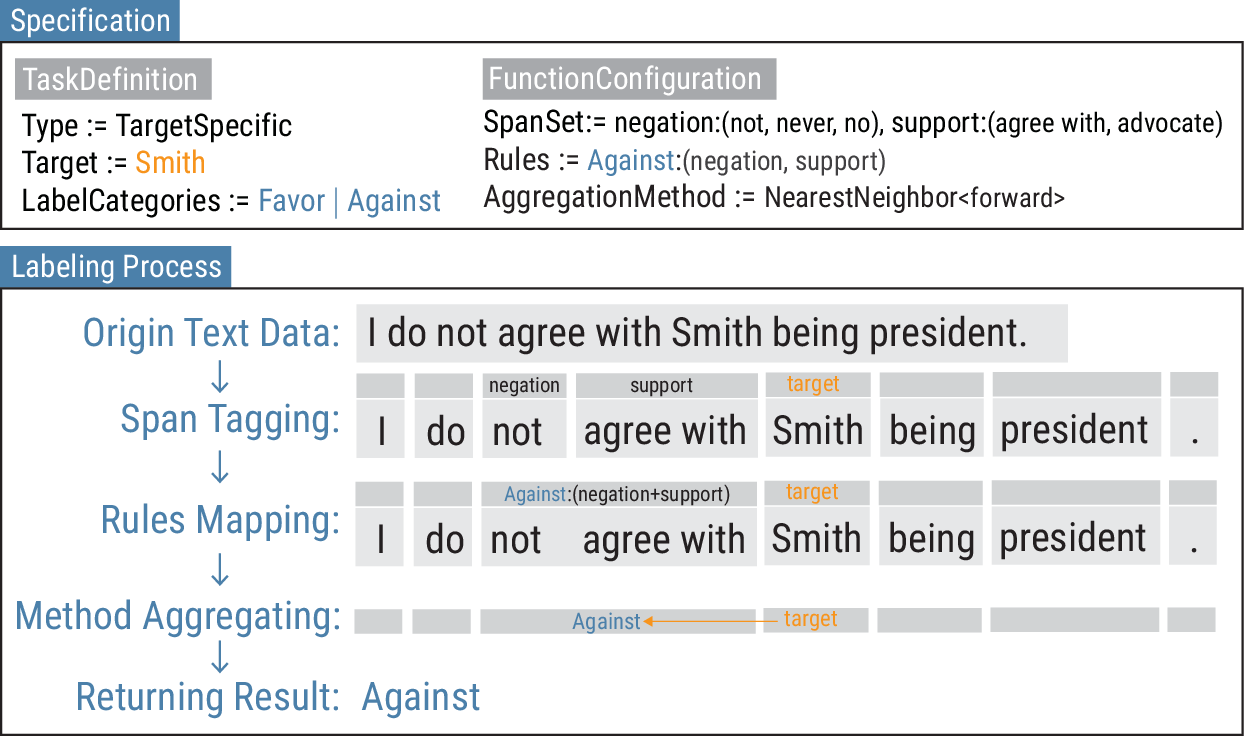}
  \caption{
    An example labeling function in our specification operates in the following steps:
    (1) Tag spans with span sets.
    (2) Apply heuristic rules to convert span sets into label categories.
    (3) Execute an aggregation method to determine the label category for this data.
    (4) Return it as the final label.
  }
  \Description{
    A flow chart illustrating the four-step labeling process.
    Step one shows tagging spans within a dataset using span sets.
    Step two demonstrates applying heuristic rules to map these span sets to specific label categories.
    Step three shows an aggregation method being executed to synthesize these inputs.
    The final step shows the output being returned as a definitive label category.
  }
  \label{fig:logic}
\end{figure}

\subsubsection{Example}

We use an example stance detection task, as shown in \cref{fig:logic}, to demonstrate how to create a labeling function according to the specification and how the labeling function operates.

\begin{itemize}[leftmargin=3.5mm]
  \item \textbf{Goal}: The goal in this example task is to classify whether the stance expressed in a text is \lfSpecCategory{Favor} or \lfSpecCategory{Against} with respect to the target \lfSpecTarget{Smith}.
  \item \textbf{Input}: ``I do not agree with Smith being president.''
  \item \textbf{Output}: \lfSpecCategory{Favor} or \lfSpecCategory{Against}
\end{itemize}

Based on the goal, we define the \lfSpecSymbol{Type} of the task as \lfSpecType{TargetSpecific}, since it aims to detect the stance toward a particular entity mentioned in the text rather than the overall sentiment of the entire passage.
The \lfSpecSymbol{Target} entity is set as \lfSpecTarget{Smith}, and the \lfSpecSymbol{LabelCategories} are defined as \lfSpecCategory{Favor} and \lfSpecCategory{Against}.
Consequently, labeling functions constructed under this configuration process only those instances that mention \lfSpecTarget{Smith}, attempting to assign a label to this entity.

In this example, we observe that the span ``agree with'' conveys a supportive meaning.
Therefore, it can be added to the \lfSpecSymbol{SpanSet} associated with the concept of \textit{support} to represent supportive expressions.
Additional spans with similar meanings, such as ``trust'', ``believe'', ``agree with'', and ``back up'', can also be added to this span set.
We can then define a \lfSpecSymbol{Rule} $\{Favor: \textit{support}\}$ so that a supportive meaning is reflected in the label category \lfSpecCategory{Favor}.

Furthermore, the word ``not'' reverses the polarity of support, and when combined with ``agree with'' expresses a negative attitude.
Using these span sets, we can tag ``not'' with the \textit{negation} span set and define a \lfSpecSymbol{Rule} $\{Against:(negation, support)\}$.
This rule indicates that when a \textit{negation} span precedes a supportive span, the combination expresses an \lfSpecCategory{Against} stance.
This captures the polarity-reversing effect of negation when paired with a supportive cue, enabling the text to be correctly classified as \lfSpecCategory{Against}.

Finally, we observe that spans expressing a \textit{Against} stance appear before the target \lfSpecTarget{Smith}.
Therefore, we adopt the \lfSpecAggregationMethod{NearestNeighbor} method and set the direction to \textit{forward} to capture this linguistic phenomenon and determine the final label category for the target.

We employ Ratner et al.'s method \cite{Ratner2019Training} to aggregate the outputs of multiple labeling functions and infer a single label category for each data sample.
The span sets and labeling functions generated for one task can be seamlessly transferred to subsequent tasks with similar labeling scenarios, thereby enhancing reusability and reducing redundant effort.

\subsection{Active Learning Module}
\label{sec:active-learning-module}

Labels generated by labeling functions are often noisy and require further refinement.
DALL uses active learning to identify instances that are most informative for review \textbf{(R2)}, aligning with prior work that employs active learning to refine noisy labels (see \cref{sec:active-learning}).
DALL provides three active learning strategies: model uncertainty sampling (\cref{sec:model-uncertainty-sampling}), query-by-committee of labeling functions (\cref{sec:query-by-committee-over-labeling-functions}), and abstention sampling (\cref{sec:abstain-sampling}).
Instances selected by these strategies are then analyzed by the LLM module to assist users in correcting labels and refining labeling functions (\cref{sec:llm-module}).

\subsubsection{Model Uncertainty Sampling}
\label{sec:model-uncertainty-sampling}

DALL uses the uncertainty-based sampling approach~\cite{Lewis1994Heterogeneous}, and specifically the margin sampling method~\cite{Scheffer2001Active}.
DALL trains a probabilistic model on data labeled by the labeling functions and predicts label probabilities for each data instance.
For each instance, margin sampling examines the difference between the top-1 and top-2 predicted class probabilities.
A smaller margin indicates less confidence in distinguishing the two leading predictions.
The margin for instance $x$ is defined as:
$$\text{margin}(x) = P(y_1 | x; \theta) - P(y_2 | x; \theta)$$
\begin{itemize}[leftmargin=3.5mm]
  \item $\theta$: the probabilistic model parameters.
  \item $y_1$, $y_2$: the first and second most likely label categories for $x$ under the model $\theta$.
  \item $P(y_i | x; \theta)$: the probability of $y_i$ for $x$ under the model $\theta$.
\end{itemize}
In our practice, we use RoBERTa~\cite{Liu2019RoBERTa} as the probabilistic model, and select the 10\% of data instances with the smallest margin for manual or LLM-assisted review.

\subsubsection{Query-by-Committee over Labeling Functions}
\label{sec:query-by-committee-over-labeling-functions}

Different labeling functions may produce conflicting labels for the same instance.
Such conflicts indicate instances that are valuable to review.
DALL adopts the query-by-committee approach~\cite{Seung1992Query} and instantiates it by treating labeling functions as the committee.
We use the vote entropy method~\cite{Dagan1995CommitteeBased} to measure disagreement.
The vote entropy for instance $x$ is defined as:
$$\text{vote\_entropy}(x) = -\sum_{i}{\frac{V(y_i, x)}{|C|}\log\frac{V(y_i, x)}{|C|}}$$
\begin{itemize}[leftmargin=3.5mm]
  \item $y_i$: a label in the set of possible labels.
  \item $V(y_i, x)$: the number of labeling functions that predict $y_i$ for $x$.
  \item $|C|$: the committee size, i.e., the number of labeling functions.
\end{itemize}
Higher vote entropy indicates greater disagreement among the labeling functions.
In our practice, we select the 10\% with the highest vote entropy for review.

\subsubsection{Abstain Sampling}
\label{sec:abstain-sampling}

In data programming~\cite{Ratner2017Snorkel}, ``abstained instances'' refer to data instances that receive no label from any labeling function.
In other words, they exhibit characteristics not yet captured by the existing labeling functions.
Reviewing such instances helps annotators discover gaps and enrich labeling functions to improve coverage.
DALL therefore provides an abstain sampling strategy that selects all abstained instances for review.

\subsection{LLM Module}
\label{sec:llm-module}

\begin{figure}[h]
  \includegraphics[width=\linewidth]{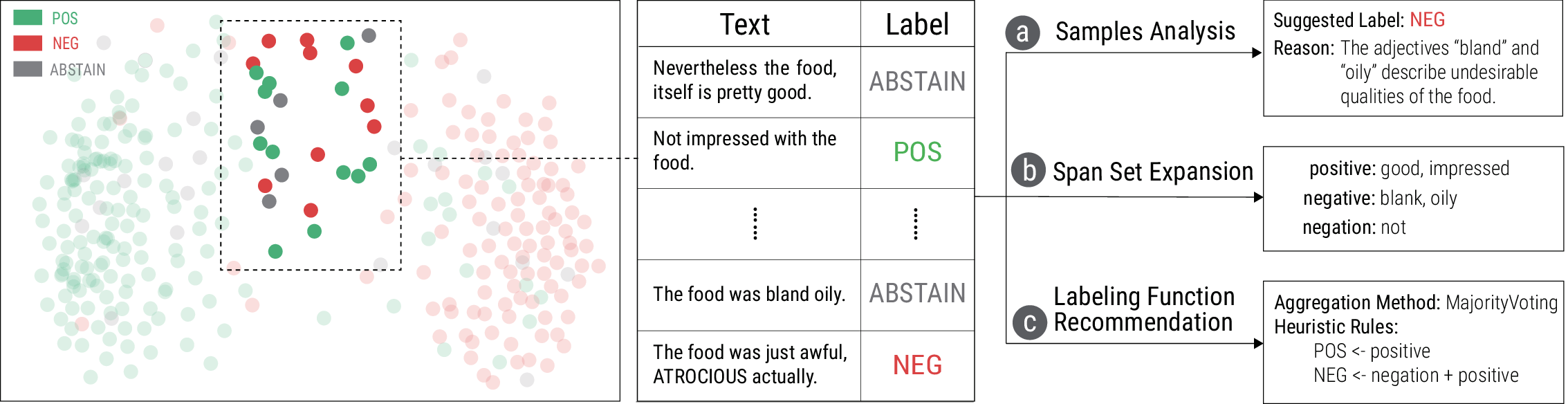}
  \caption{
    The LLM module supports the labeling process in the following ways:
    (a) Expand span sets using spans extracted from samples.
    (b) Recommend labeling functions from generalizable heuristics.
    (c) Analyze samples and provide label recommendations for user reference.
  }
  \Description{
    A diagram showcasing the LLM analysis module's three main components.
    Component A shows extracting spans from data samples to build span sets.
    Component B illustrates examining samples to generate heuristic rules and labeling function recommendations.
    Component C displays providing direct label recommendations based on sample analysis.
  }
  \label{fig:llm}
\end{figure}

DALL uses LLMs to assist the labeling process in three ways (see \cref{fig:llm}), all operating on samples selected by the active learning module (\cref{sec:active-learning-module}):
\begin{itemize}[leftmargin=3.5mm]
  \item \textbf{Sample Analysis}: analyze samples and provide label recommendations.
  \item \textbf{Span Set Expansion}: expand span sets using spans extracted from samples.
  \item \textbf{Labeling Function Recommendation}: recommend labeling functions from generalizable heuristics.
\end{itemize}

\begin{figure}
  \includegraphics[width=0.9\linewidth]{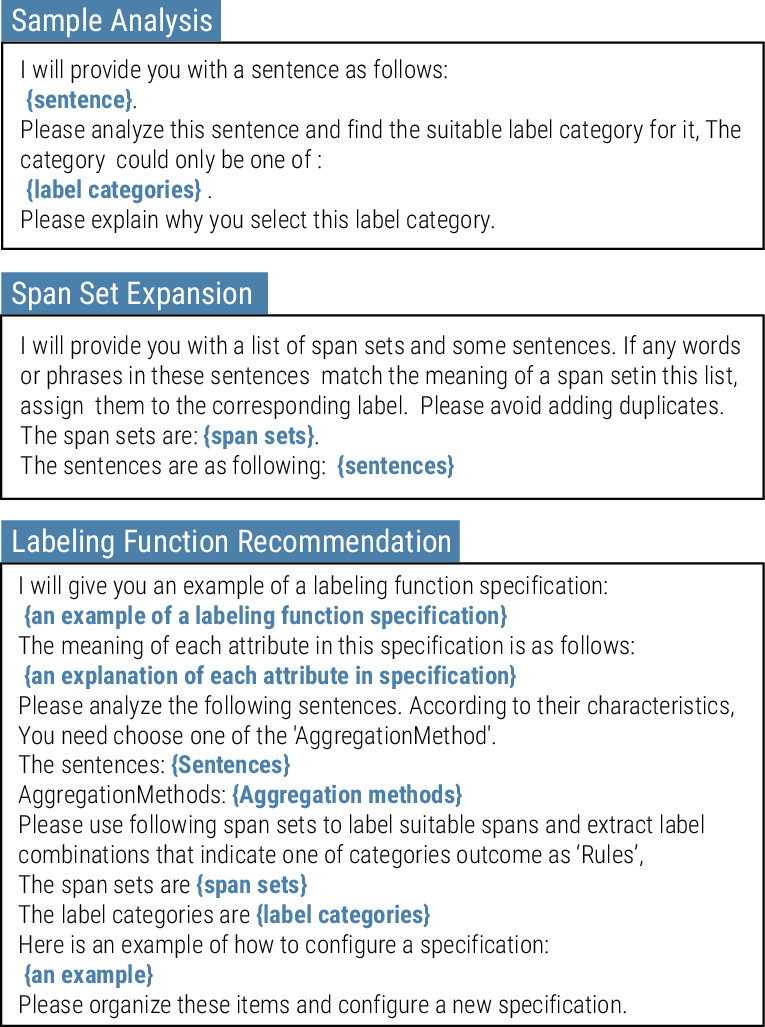}
  \caption{
    The prompts for the three components of the LLM module: sample analysis, span set expansion, and labeling function recommendation. Text in blue indicates variables whose values depend on the current labeling status.
  }
  \Description{
    The figure displays three structured text templates for LLM prompts, corresponding to sample analysis, span set expansion, and labeling function recommendation.
    Each template contains static instruction text and dynamic variables highlighted in blue.
    These blue segments indicate variables to be injected into the prompt.
  }
  \label{fig:llmprompt}
\end{figure}

Span set expansion and labeling function recommendation guide refinement of labels (\textbf{R3}).
Sample analysis provides label recommendations for user reference (\textbf{R2}).
The prompt templates for these three components are shown in \cref{fig:llmprompt}.
The following describes the details for the three components.

\textbf{Sample Analysis.}
For instances selected by active learning, DALL uses the LLM to generate recommended labels with detailed rationales.
This leverages the contextual reasoning of LLMs for cases that are difficult for span-level labeling functions (e.g., sarcasm or idiomatic expressions), thereby supporting more accurate label correction.
The rationales may also guide users to design new heuristic rules and refine existing labeling functions.

\textbf{Span Set Expansion.}
The effectiveness of labeling functions in DALL relies on the \lfSpecSymbol{SpanSet} (\cref{sec:function-configuration}), yet manually specifying span sets through span tagging is laborious.
DALL uses the LLM to analyze selected samples and identify spans relevant to each span set.
Users provide a few representative examples by tagging a few spans per set.
The LLM infers the intended semantics and then detects relevant spans in the samples selected by active learning.

\textbf{Labeling Function Recommendation.}
To improve labeling functions, DALL analyzes the selected samples to extract generalizable heuristic rules, chooses suitable aggregation methods for the task, and recommends new or refined labeling functions.
Recommendations are given as structured specifications described in \cref{sec:data-programming-module} in JSON format, which are converted into executable code.

\section{DALL Labeling System}

Based on the DALL framework introduced in \cref{sec:framework}, we developed the DALL labeling system for text labeling tasks.

\begin{figure*}
  \centering
  \includegraphics[width=\textwidth]{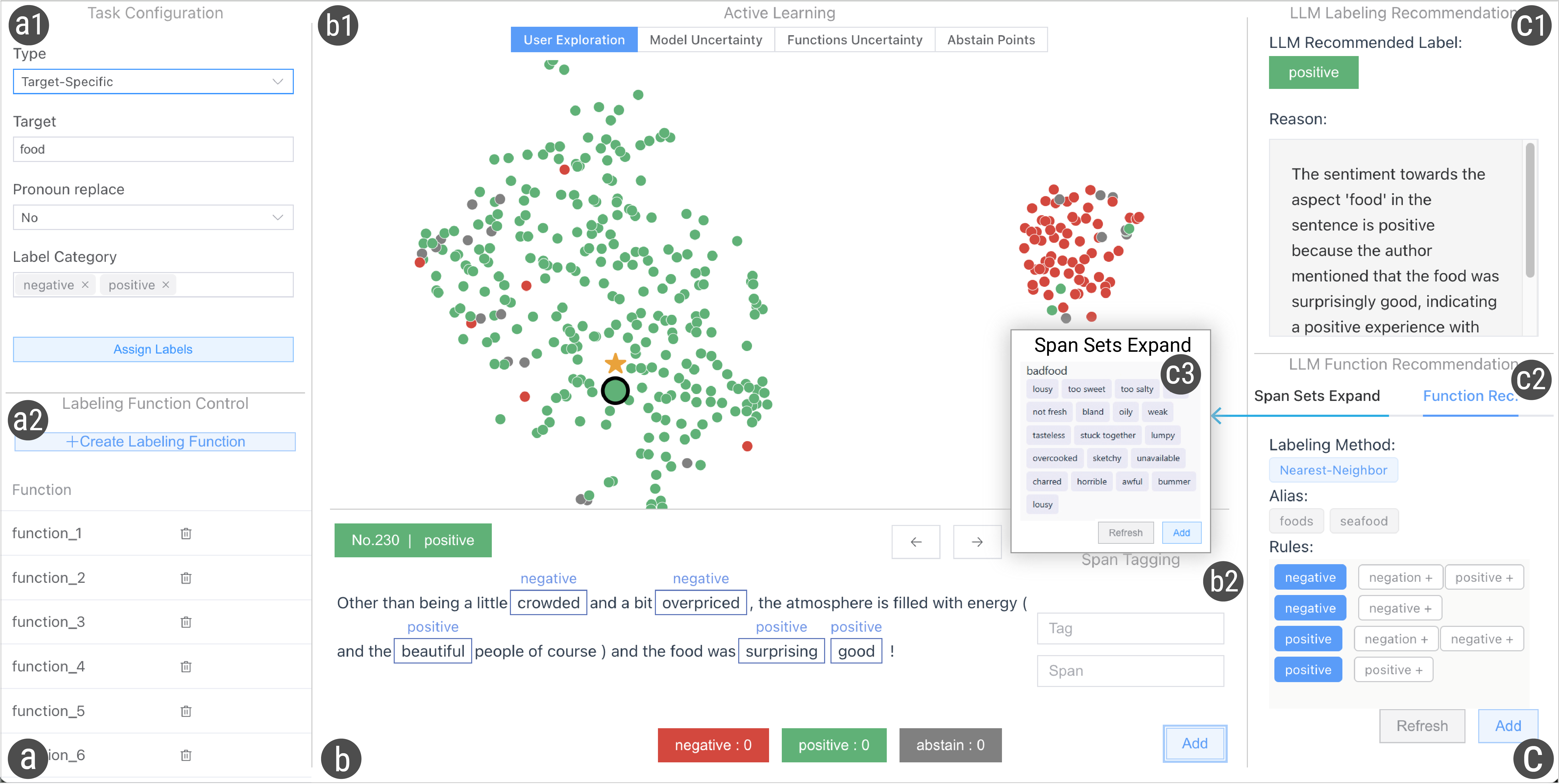}
  \caption{
    The system interface consists of three panels:
    (a) The Labeling Function Panel allows users to construct labeling functions and apply them to automatically annotate data.
    (b) The Active Learning Panel enables users to explore data under different sampling strategies, elicit heuristic rules via interactive span tagging, and verify labels.
    (c) The LLM Analysis Panel provides assistance from LLMs by analyzing instances, expanding span sets, and recommending labeling functions.
  }
  \Description{
    A screenshot of the system interface showing three main panels.
    On the left, the Labeling Function Panel (a) contains widgets for constructing and applying functions.
    In the center, the Active Learning Panel (b) displays data instances for interactive span tagging and label verification.
    On the right, the LLM Analysis Panel (c) provides AI-generated analysis, span suggestions, and function recommendations.
    The panels are visually integrated to support a seamless labeling workflow.
  }
  \label{fig:system}
\end{figure*}

\subsection{User Interface}

The interface (\cref{fig:system}) comprises three main panels aligned with the DALL framework:
\begin{itemize}[leftmargin=3.5mm]
  \item \textbf{Labeling Function Panel} for data programming (\cref{fig:system}(a)).
  \item \textbf{Active Learning Panel} for exploring and refining labels (\cref{fig:system}(b)).
  \item \textbf{LLM Analysis Panel} for LLM assistance (\cref{fig:system}(c)).
\end{itemize}
We describe each panel in detail below.

\subsubsection{Labeling Function Panel}

Within the labeling function panel, the task configuration panel (\cref{fig:system}(a1)) and the labeling function control panel (\cref{fig:system}(a2)) support creating and applying labeling functions.

\textbf{Task configuration panel} implements the task definition component of the specification (see \cref{fig:system}(a1)).
It allows users to specify the \lfSpecSymbol{TaskDefinition} for the labeling task, with \lfSpecSymbol{Type}, \lfSpecSymbol{Target}, and \lfSpecSymbol{LabelCategories}, according to the specification in \cref{sec:task-definition}.
The task definition defines the scope within which one or more labeling functions operate.
Through this panel, users can trigger label assignment by clicking the ``Assign Labels'' button.
It aggregates all labeling functions and assigns labels to the dataset as described in \cref{sec:data-programming-module}, inferring a single consensus label per instance.

\begin{figure*}
  \centering
  \includegraphics[width=0.95\textwidth]{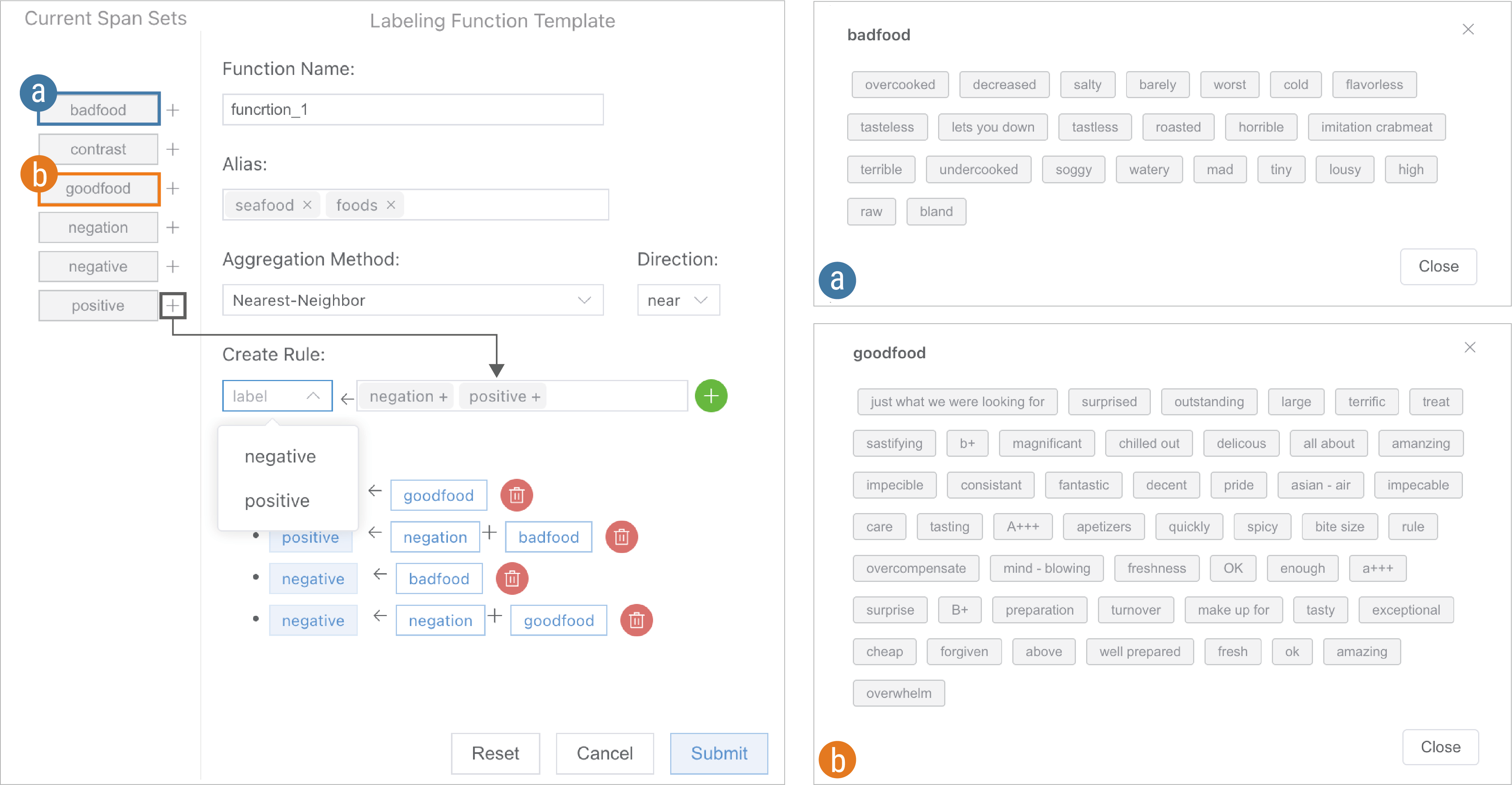}
  \caption{
    The Labeling Function Construction Panel.
    This panel implements the specification introduced in \cref{sec:data-programming-module} and is designed to help users create effective and executable labeling functions through direct selection and input.
  }
  \Description{
    A screenshot showing the construction interface for labeling functions.
    The interface features dropdown menus, selection boxes, and input fields for task definitions and function configurations.
    The interface uses form-based elements that allow users to build logic through visual selection and text input, directly implementing the framework's formal specification.
  }
  \label{fig:construct}
\end{figure*}

\textbf{Labeling function control panel} implements the function configuration component of the specification (see \cref{fig:system}(a2)).
It allows users to create and manage labeling functions, which are displayed as a list.
Clicking the ``Create Labeling Function'' button opens the labeling function construction panel (see \cref{fig:construct}), where the user defines a function name.
For target-specific tasks, the user can assign multiple aliases to the target so that all occurrences of those aliases in the text are recognized as the same target.
The construction panel lists current span sets on the left; the user can expand each set to inspect spans.
The user combines span sets with the ``+'' button and assigns a label category to form rules, which appear in the lower part of the panel.
After clicking the ``Submit'' button, the new labeling function is added to the list in the labeling function control panel.

\subsubsection{Active Learning Panel}

Within the active learning panel, the distribution panel (\cref{fig:system}(b1)) visualizes instances and active learning results, and the instance inspection panel (\cref{fig:system}(b2)) shows the selected instance and supports span tagging and label verification.

\textbf{Distribution panel} enables users to visually assess labeling status and active learning outcomes (see \cref{fig:system}(b1)).
Within the scatter plot, each point represents a piece of text.
The scatter plot is generated by PCA projection of the text embeddings computed with RoBERTa.
Point color encodes the assigned label.
Users can switch between the active learning strategies described in \cref{sec:active-learning-module}.
When an active learning strategy is selected, the corresponding sampled instances are highlighted.

\textbf{Instance inspection panel} provides access to a selected instance to inspect text, tag spans, and verify or correct labels (see \cref{fig:system}(b2)).
Users can manually tag spans.
This panel also serves the purpose of showing the raw data to inspire the user to identify patterns for creating labeling functions.
Navigation between instances can be done via switch buttons (``←'' and ``→'') or by selecting a point in the distribution panel.

\subsubsection{LLM Analysis Panel}

Within the LLM analysis panel, the labeling recommendation panel (\cref{fig:system}(c1)) provides label suggestions and rationales for selected instances, and the labeling function recommendation panel (\cref{fig:system}(c2)) supports span set expansion and labeling function recommendations.

\textbf{Labeling recommendation panel} shows the LLM-generated label suggestions and rationales for selected instances (see \cref{fig:system}(c1)).
It corresponds to the sample analysis component of the LLM module (\cref{sec:llm-module}).
It helps users verify labels and gain insights for refining labeling functions.

\textbf{Labeling function recommendation panel} shows the LLM-generated span set expansion (see \cref{fig:system}(c3)) and labeling function recommendations for selected instances (see \cref{fig:system}(c2)).
It corresponds to the span set expansion and labeling function recommendation components of the LLM module (\cref{sec:llm-module}).
It consists of two parts:
\begin{itemize}[leftmargin=3.5mm]
  \item The ``Span Set Expansion'' part displays spans from samples selected by active learning that match each existing span set.
  \item The ``Function Recommendation'' part (labeled ``Function Rec.'' in \cref{fig:system}(c)) recommends labeling functions based on these samples.
\end{itemize}

\subsection{User Interaction}

\begin{figure*}
  \centering
  \includegraphics[width=0.95\textwidth]{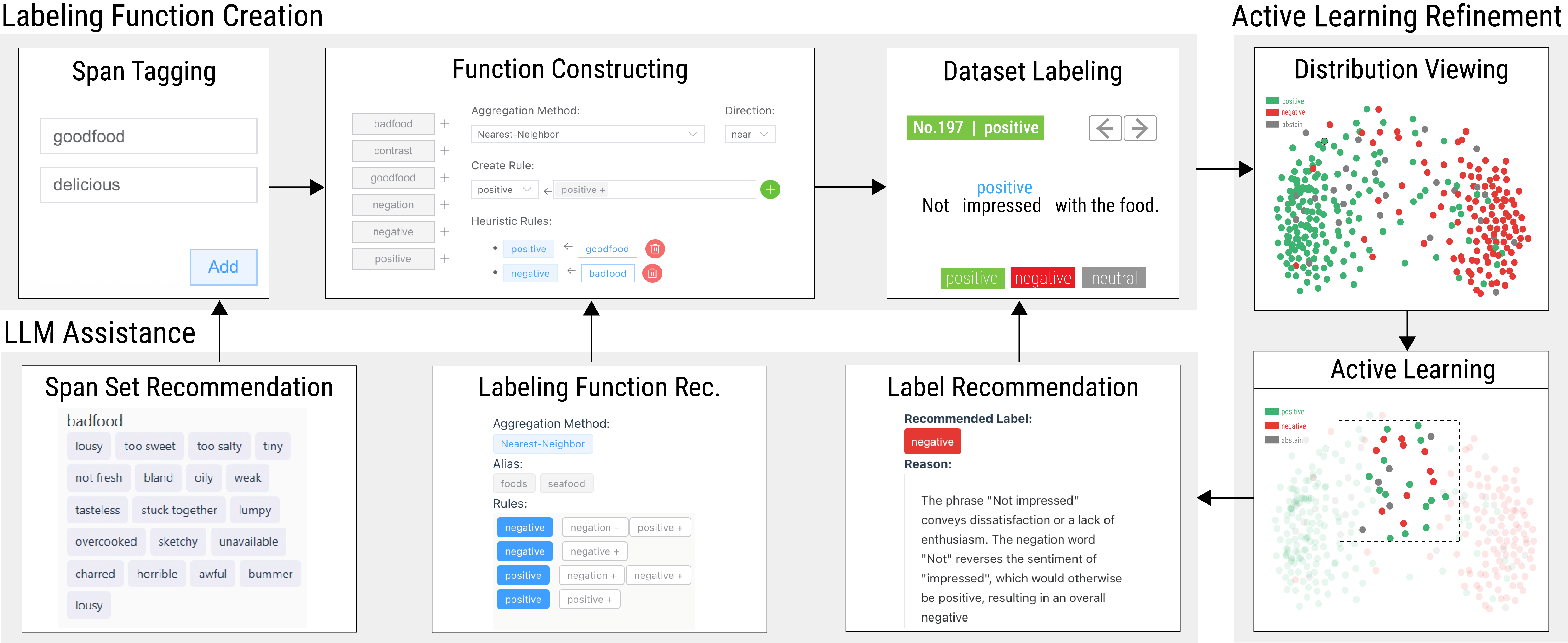}
  \caption{
    The usage workflow of the DALL labeling system:
    (1) The user tags a few spans and creates initial labeling functions, then assigns labels to the dataset.
    (2) The distribution panel updates.
    The user applies active learning strategies and selects highlighted instances to inspect.
    (3) The LLM analyzes selected instances and provides label recommendations, span suggestions, and labeling function recommendations.
    The user can incorporate these into refined labels and functions.
    This workflow can be repeated to progressively improve the labeling quality.
  }
  \Description{
    A circular workflow diagram consisting of three main stages.
    Stage one shows the initial user input through manual span tagging and basic labeling function creation.
    Stage two illustrates a feedback loop where the dataset distribution view updates, and the user selects informative samples via active learning strategies.
    Stage three displays the LLM analysis of these samples, providing recommendations that the user incorporates back into the labeling functions.
    Arrows connect these stages to indicate a continuous, iterative cycle for improving labeling quality.
  }
  \label{fig:workflow}
\end{figure*}

The system supports an interactive workflow that mirrors the DALL framework (\cref{sec:framework}).
The user creates and applies labeling functions (data programming), uses active learning to focus on informative instances, and draws on LLM analysis to correct labels and refine labeling functions, as shown in \cref{fig:workflow}.
This process can be repeated to progressively improve the labeling quality.

\textbf{Labeling function creation.}
The process starts with data exploration in the instance inspection panel or by selecting points in the distribution panel (\cref{fig:system}(b1) and (b2)).
The user derives heuristics from instance content and turns them into labeling functions via the labeling function construction panel (\cref{fig:construct}).
Specifically, the user chooses span sets, forms rules by combining span sets and assigning label categories, and selects an aggregation method (\cref{sec:function-configuration}).
After creating one or more labeling functions, the user clicks the ``Assign Labels'' button in the task configuration panel to run the data programming pipeline (\cref{sec:data-programming-module}), which assigns a consensus label to each instance and updates the distribution view.

\textbf{Active learning refinement.}
After the label assignment, the distribution panel reflects the current labels and supports the active learning strategies in \cref{sec:active-learning-module}.
Selecting a strategy highlights the corresponding instances in the scatter plot.
The user can click a point to view the instance in the instance inspection panel, where the user can correct labels, tag spans, or derive new heuristics and update labeling functions.

\textbf{LLM assistance.}
The LLM provides assistance in three ways.
First, after the user provides a few representative spans per span set, the labeling function recommendation panel (\cref{fig:system}(c2)) uses the LLM to expand span sets.
Second, this panel also recommends new labeling functions from the samples selected by active learning, reducing manual tagging and speeding up coverage.
The user can apply suggested span sets and labeling functions with one click.
Third, for user-selected instances, the labeling recommendation panel (\cref{fig:system}(c1)) provides LLM-generated label recommendations and rationales, assisting the user in correcting labels and inspiring new rules for labeling functions.

\section{Evaluation}
\label{sec:evaluation}

We conducted three studies: a comparative study of our system against existing labeling systems, an ablation study of DALL's three core modules, and a usability study.

\subsection{Comparative Study}
\label{sec:comparative-study}

We conducted a comparative study to evaluate the effectiveness of the DALL labeling system against existing labeling systems.

\subsubsection{Study Design}

We compared the DALL labeling system with three alternatives: \evaluatedMethodDoccano~\cite{Nakayama2018doccano}, \evaluatedMethodSnorkel~\cite{Ratner2017Snorkel}, and LLM-based labeling with GPT-3.5 Turbo (referred to as \evaluatedMethodGPT)~\cite{OpenAI2023GPT35}.
\evaluatedMethodGPT is automated and does not involve participants.

\textbf{Participants.}
Thirty-six participants (referred to as P1--P36) aged between 22 and 40 from local universities and industry took part in the study.
All participants had computer science backgrounds and data processing experience.

\textbf{Apparatus.}
The study was run on a workstation with an NVIDIA RTX A6000 GPU (48 GB GPU memory) and a 23-inch 1920$\times$1080 LCD monitor.

\textbf{Procedure.}
Participants were randomly assigned to four groups: \evaluatedMethodDoccano (P1--P9), \evaluatedMethodSnorkel (P10--P18), \evaluatedMethodDALL (P19--P27), and \evaluatedMethodReusedDALL (P28--P36).
All groups received a short tutorial on their system and were instructed to balance labeling efficiency and accuracy.
All participants completed two stages.
Both stages involved labeling sentiment analysis datasets, so that labeling functions from the first stage could be reused in the second to evaluate reusability of the DALL system.

\textbf{Stage 1} was a target-specific classification task on a small dataset.
The dataset contains 370 restaurant reviews from SemEval-2014 Task 4~\cite{Pontiki2014SemEval2014}, with aspect-level sentiment labels (positive/negative) toward food.
P1--P36 used their assigned system.
Only the results of P1--P27 were used in the comparative analysis (i.e., the analysis described in \cref{sec:comparative-study-results} for the target-specific classification task).
We recorded labeling accuracy at 5-minute intervals over the first 30 minutes (unlabeled instances counted as mislabeled), the final accuracy, and total task time.

\textbf{Stage 2} was a text classification task on a large dataset.
The dataset contains 10,000 instances randomly sampled from Sentiment140~\cite{Go2009Twitter} to assess scalability of the system and reusability of the created labeling functions.
P1--P36 continued with the same system.
For P1--P27, their labeling functions created in stage 1 were removed and they started from scratch.
For P28--P36, their labeling functions created in stage 1 were kept and could be applied to the new dataset (we refer to this condition as \evaluatedMethodReusedDALL).
We tracked accuracy over the first 30 minutes.
As fully labeling 10,000 instances was infeasible within the study period, we also recorded the time and resources needed for each system to reach 85\% accuracy as a measure of efficiency at scale.

\begin{figure}
  \centering
  \includegraphics[width=\linewidth]{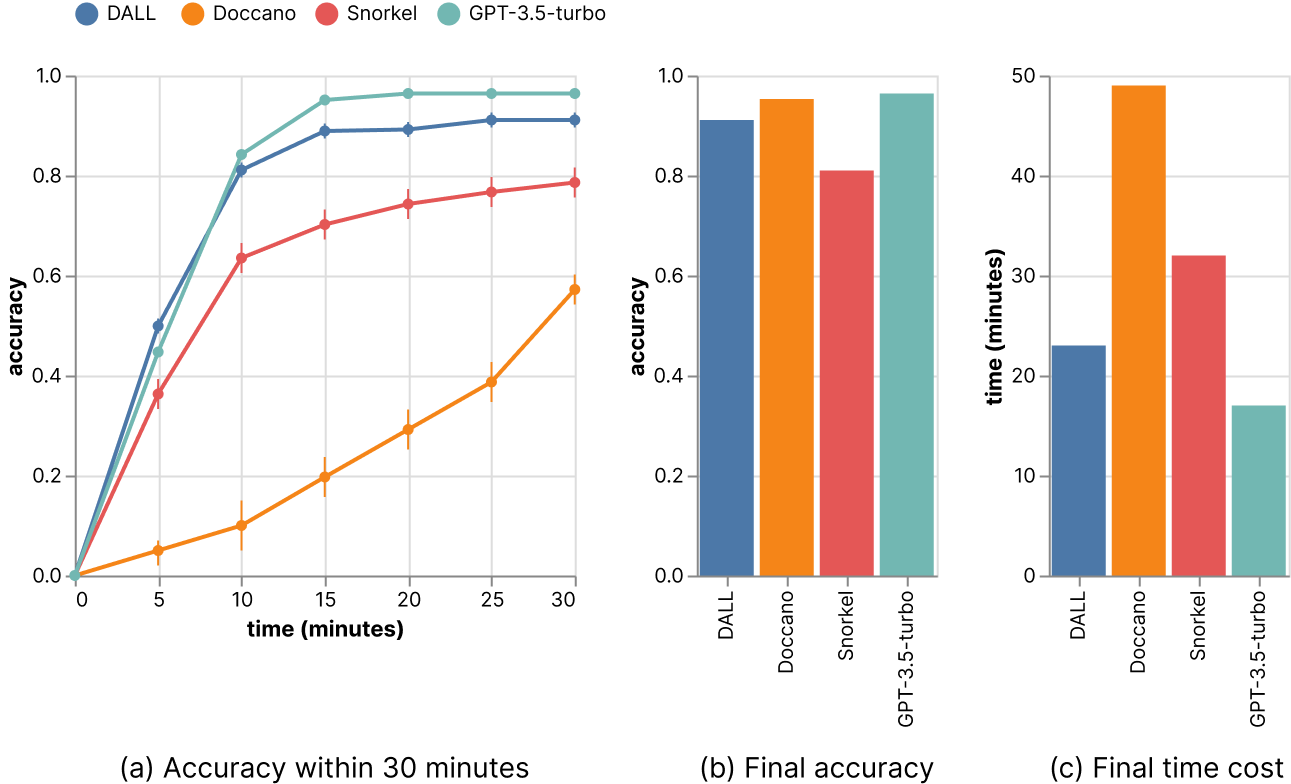}
  \caption{
    Comparative study result for the target-specific task (stage 1).
    (a) Accuracy within 30 minutes. %
    (b) Final accuracy.
    (c) Final time cost.
  }
  \Description{
    Comparative study result for the target-specific task (stage 1).
    Subfigure (a) is a line chart showing labeling accuracy over time.
    Subfigure (b) is a bar chart showing final labeling accuracy by method.
    Subfigure (c) is a bar chart showing total labeling time by method.
  }
  \label{fig:comparative-study-target-specific}
\end{figure}

\subsubsection{Results}
\label{sec:comparative-study-results}

The following describes the results for the target-specific classification task (stage 1) and the text classification task (stage 2).

\textbf{Target-specific classification task (stage 1).}
\Cref{fig:comparative-study-target-specific}(a) shows that the label accuracy of \evaluatedMethodGPT and \evaluatedMethodDoccano improves roughly linearly over time.
\evaluatedMethodSnorkel and \evaluatedMethodDALL both show sharp initial improvements with the addition of labeling functions.
While \evaluatedMethodDALL maintains a steadier trajectory, \evaluatedMethodSnorkel's progress slows later on, possibly due to overlapping logic or diminishing marginal utility of additional functions.
\evaluatedMethodDALL sustains improvement by using active learning to surface ambiguous instances and LLM-guided refinement to correct labels and refine labeling functions.
\Cref{fig:comparative-study-target-specific}(b) and (c) show that on this small dataset \evaluatedMethodDALL reaches final accuracy comparable to \evaluatedMethodGPT and \evaluatedMethodDoccano at a competitive time cost, and clearly outperforms \evaluatedMethodSnorkel, with time cost slightly above \evaluatedMethodGPT.

\begin{figure}
  \centering
  \includegraphics[width=\linewidth]{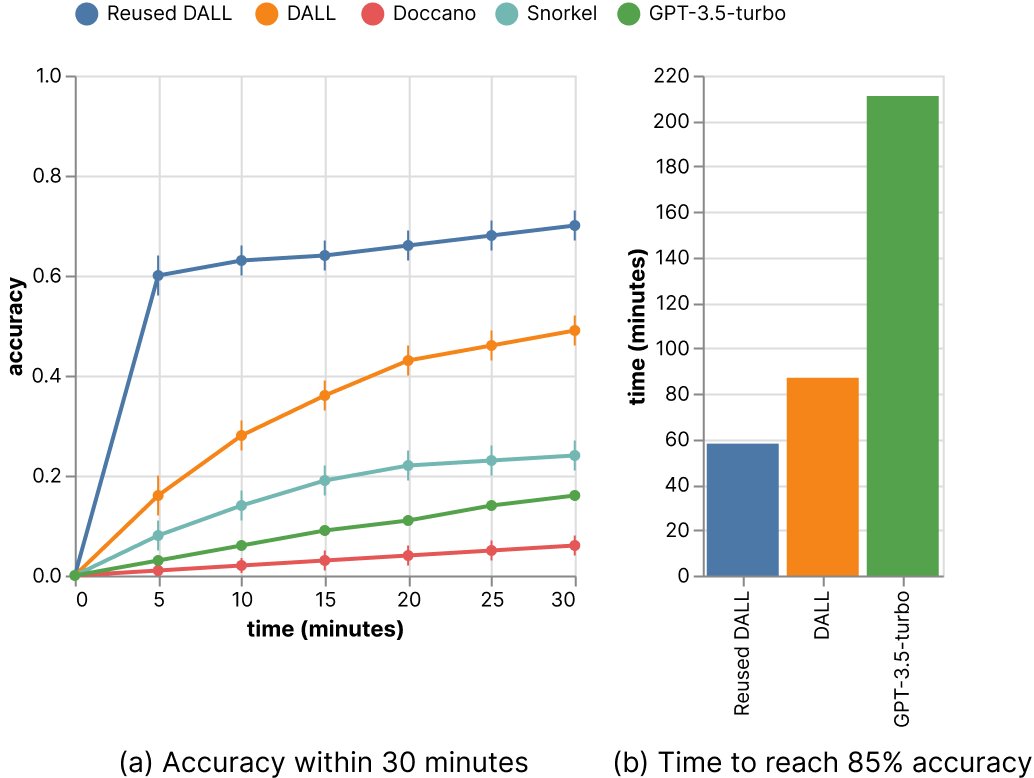}
  \caption{
    Comparative study result for the text classification task (stage 2).
    (a) Accuracy within 30 minutes.
    (b) Time to reach 85\% accuracy.
}
  \Description{
    This figure shows the results of the comparative study for the text classification task.
    Subfigure (a) is a line graph showing labeling accuracy over the first 30 minutes.
    Subfigure (b) is a bar chart showing time to reach 85\% accuracy by method.
  }
  \label{fig:comparative-study-text-classification}
\end{figure}

\textbf{Text classification task (stage 2).}
\Cref{fig:comparative-study-text-classification} shows that \evaluatedMethodDALL reaches comparable accuracy with substantially less labeling time.
Reusing labeling functions from stage 1 (\evaluatedMethodReusedDALL) allowed rapid convergence to high accuracy over time (\cref{fig:comparative-study-text-classification}(a)).
We did not record the total labeling time for fully labeling the 10,000-instance dataset because of the prohibitive cost.
Instead, we measured the time cost to reach 85\% accuracy (\cref{fig:comparative-study-text-classification}(b)).
\evaluatedMethodDALL reached this level of accuracy significantly faster and with lower resource use.
It offers a better trade-off between accuracy and time cost.
\evaluatedMethodGPT's time cost grows linearly with data volume and can be slow for large datasets.

\subsection{Ablation Study}

We conducted an ablation study to evaluate whether each of the three modules described in \cref{sec:framework}, data programming, active learning, and LLM analysis, contributes to efficiency and label quality.

\subsubsection{Study Design}

We compared three configurations, incrementally adding modules:

\begin{itemize}[leftmargin=3.5mm]
  \item \textbf{Data Programming (\evaluatedMethodDP)}: The participant can use data programming but cannot use active learning and LLM analysis.
    In this setup, the distribution panel (\cref{fig:system}(b1)) and LLM analysis panel (\cref{fig:system}(c)) were disabled.
  
  \item \textbf{Data Programming + Active Learning (\evaluatedMethodDPAL)}: The participant can use data programming and active learning but cannot use LLM analysis.
    In this setup, the LLM analysis panel (\cref{fig:system}(c)) was disabled.
  
  \item \textbf{Data Programming + Active Learning + LLM (\evaluatedMethodDPALLLM)}: The participant can use data programming, active learning, and LLM analysis.
    In this setup, all the functionalities of the system (\cref{fig:system}) are available.
\end{itemize}

\textbf{Participants.}
Eighteen participants (P1--P18) from the comparative study who had not used the DALL system were recruited to avoid familiarity bias.
They were randomly assigned to one of the three configurations (\evaluatedMethodDP, \evaluatedMethodDPAL, or \evaluatedMethodDPALLLM).

\textbf{Dataset and Procedure.}
We used the AG News dataset~\cite{Zhang2015Characterlevel}, a text classification benchmark with news articles in four categories: \lfSpecCategory{World}, \lfSpecCategory{Sports}, \lfSpecCategory{Business}, and \lfSpecCategory{Sci/Tech}.
The procedure followed that of the comparative study (\cref{sec:comparative-study}).

\begin{figure}
  \centering
  \includegraphics[width=\linewidth]{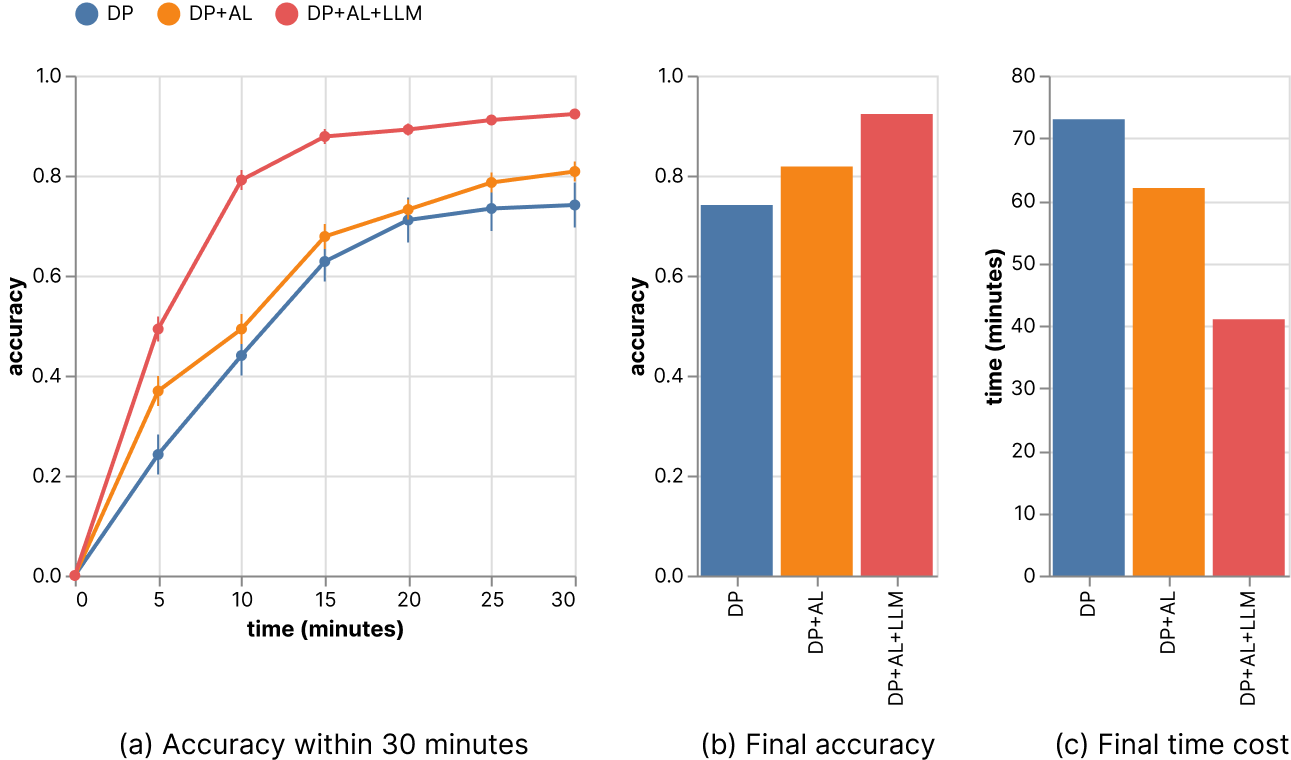}
  \caption{
    Ablation study results.
    (a) Accuracy within 30 minutes.
    (b) Final accuracy.
    (c) Final time cost.
  }
  \Description{
    This figure shows the results of the ablation study, comparing the configurations of DP, DP+AL, and DP+AL+LLM.
    Subfigure (a) is a line chart showing labeling accuracy over time.
    Subfigure (b) is a bar chart showing final labeling accuracy.
    Subfigure (c) is a bar chart showing final labeling time cost.
  }
  \label{fig:ablation-study}
\end{figure}

\begin{figure*}
  \centering
  \includegraphics[width=0.92\textwidth]{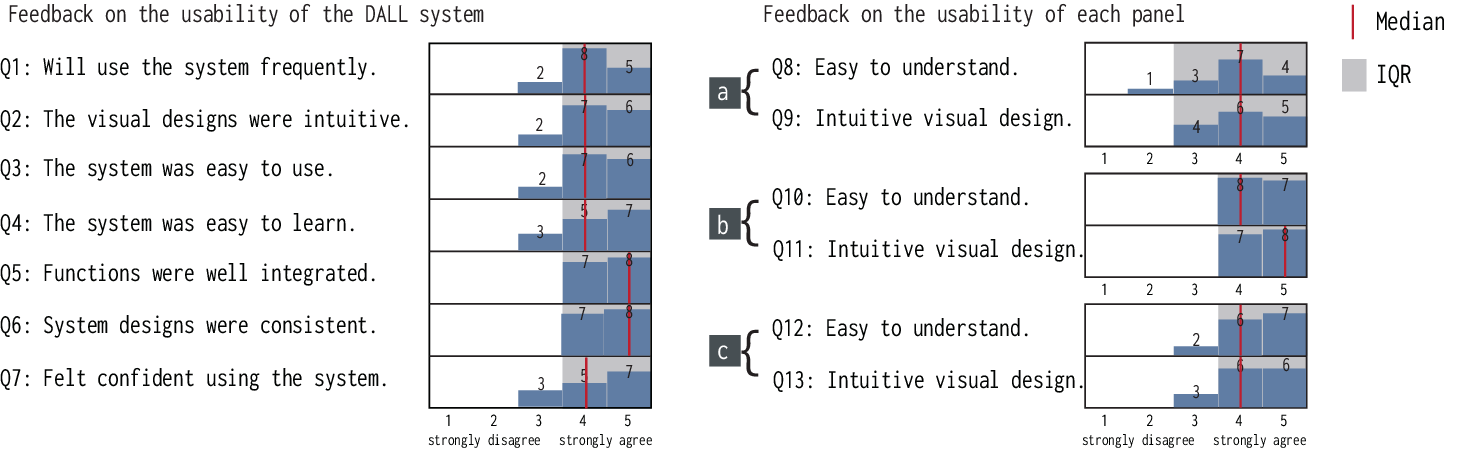}
  \caption{
    Usability study feedback.
    Left: Overall usability ratings for the DALL labeling system.
    Right: Ratings by each panel of the system:
    (a) Labeling function panel.
    (b) Active learning panel.
    (c) LLM analysis panel.
  }
  \Description{
    A collection of bar charts displaying user study feedback.
    The left side consists of a bar chart showing overall usability ratings for DALL system, such as ease of use and satisfaction.
    The right side is divided into three smaller bar charts, labeled (a), (b), and (c), which present specific user ratings for the Labeling Function Panel, the Active Learning Panel, and the LLM Analysis Panel respectively.
    Most bars indicate positive feedback with high scores across all evaluated categories.
  }
  \label{fig:rate}
\end{figure*}

\subsubsection{Results}

\Cref{fig:ablation-study} shows that labeling accuracy and efficiency improve with each added module.

\begin{itemize}[leftmargin=3.5mm]
  \item \evaluatedMethodDP:
        Without the distribution or LLM panels, participants could only browse instances sequentially or by arbitrary selection.
        Accuracy growth slowed over time and plateaued.
        Participants reported that the system often surfaced instances already covered by existing labeling functions, which made it inefficient to extend coverage to new cases.

  \item \evaluatedMethodDPAL:
        Adding active learning raised accuracy above \evaluatedMethodDP.
        Participants reported using active learning to focus on ambiguous or conflicting instances and then correcting labels or refining labeling functions based on those samples.

  \item \evaluatedMethodDPALLLM:
        Adding LLM analysis yielded a significant gain in accuracy and reduction in time cost.
        Participants reported that they could often follow the LLM's suggestions to refine labeling functions and correct labels in batch, reducing manual effort.
\end{itemize}

\subsection{Usability Study}

We collected feedback from participants who had used the full DALL system: nine from the comparative study (P19--P27) and six from the \evaluatedMethodDPALLLM group in the ablation study.
Each participant completed a lightly edited version of the System Usability Scale (SUS)~\cite{Brooke1996SUS} and a 20-minute one-on-one interview.
In the edited SUS questionnaire, we let the participant rate each of the three main panels of the system, labeling function panel, active learning panel, and LLM analysis panel.

\textbf{Overall usability.}
\Cref{fig:rate}(Left) summarizes the results.
Participants reported high satisfaction: accessibility items (Q2--Q4) had median scores of 4 or 5 with narrow spread, and most adapted to the labeling logic after a short tutorial.
Items on intention to use (Q1), confidence (Q7), and integration (Q5, Q6) indicate that participants saw the system as useful in real-world labeling and as a cohesive integration of the three modules.
P19 noted the value for individual developers: ``\quotedfeedback{This system provides an effective way for personal developers to generate datasets for their own specialized domains.}''
P22 said: ``\quotedfeedback{Overall, the system is well-developed and highly practical, with clear operational logic that allows users to quickly get started and ensures convenient daily use.}''

\textbf{Panel-level feedback.}
The following describes participant feedback on each of the three panels.

\begin{itemize}[leftmargin=3.5mm]
  \item \textbf{Active learning panel (\cref{fig:system}(b)).} This panel received the highest ratings.
  Participants valued how the distribution plot surfaces informative samples and makes active learning strategy outcomes visible.
  For example, P2: ``\quotedfeedback{By highlighting active learning results in the plot, the system presented black-box sampling strategies in a clear and accessible manner that was readily understood and accepted.}''

  \item \textbf{LLM analysis panel (\cref{fig:system}(c)).} Participants praised LLM assistance in span set expansion and in generating labeling functions for some challenging cases.
  P27: ``\quotedfeedback{The automatic span tagging helped me the most. The coverage of span sets determines the capacity of a labeling function, and expanding them is usually a labor-intensive task. However, with LLM, I could do this task much more easily.}''
  P3: ``\quotedfeedback{There were some cases for which it was difficult for me to classify or create an appropriate labeling function, but the suggestions from LLM helped me resolve them quickly.}''

  \item \textbf{Labeling function panel (\cref{fig:system}(a)).}
  Ratings were positive overall, with slight variation.
  A few participants (e.g., P21 and P5) found the configuration logic initially hard to understand and suggested clearer workflow guidance for new users.
  P5: ``\quotedfeedback{It took me a while to understand the configuration logic of labeling functions. I think the lack of guidance on the workflow of the labeling function may cause confusion for new users who may not fully understand the core features or the intended usage paths}''.
  Others highlighted the low programming burden compared to code-based tools.
  P4: ``\quotedfeedback{With Snorkel, I almost need to code a function for every instance, but when using the system, I only need to adjust the span combinations.}''
  P2: ``\quotedfeedback{After receiving the tutorial on how to create labeling functions, I could easily produce labeling functions in bulk with just a few clicks.}''
\end{itemize}

\section{Discussion}

We discuss design considerations, limitations, and future directions.

\textbf{Expressiveness of the specification.}
The specification (\cref{sec:data-programming-module}) lets users construct labeling functions without code by defining span sets and rules (\cref{sec:function-configuration}).
This approach can struggle with instances that require cross-sentence reasoning or implicit expressions.
Addressing this limitation is a potential direction for future work.
Meanwhile, the comparative and ablation results (\cref{sec:evaluation}) suggest the approach is sufficiently expressive for the studied tasks and that active learning plus the LLM effectively surface and support correction of remaining hard cases.

\textbf{Generalizability across domains.}
DALL supports diverse labeling tasks.
Domain-independent expressions (e.g., ``improve'' as a positive cue) can be captured by reusable span sets.
Domain-dependent ones need task-specific interpretation.
Thus, a future direction is to build a library of general and domain-specific span sets for the user to reuse.

\textbf{Prompt design and optimization.}
DALL uses LLMs to support sample analysis, span set expansion, and labeling function recommendation, each with a dedicated prompt template (\cref{fig:llmprompt}).
These templates are not exposed in the UI so that users can focus on labeling rather than prompt editing.
Meanwhile, prompt effectiveness can vary across datasets and tasks.
A future direction is to incorporate automatic or semi-automatic prompt optimization for new domains and task types.

\section{Conclusion}

We presented DALL, a text labeling framework that integrates data programming, active learning, and large language models to improve the trade-off between label quality and labeling cost.
DALL introduces a structured specification that lets users and LLMs define labeling functions via configuration rather than code, and combines it with active learning to focus refinement on informative instances and with LLM analysis to support label correction and labeling function refinement.
We implemented DALL as an interactive labeling system and evaluated it through comparative, ablation, and usability studies.
The studies show that the system achieves comparable or higher accuracy than alternatives with substantially lower labeling time, that its three modules jointly improve accuracy and efficiency, and that participants found it effective.

\section*{Acknowledgments}
We thank the anonymous reviewers for their valuable comments.
This work is supported by NSFC No. 62572054, 62302038, U2268205, Young Elite Scientists Sponsorship Program by CAST (Grant No. 2023QNRC001).

\bibliographystyle{ACM-Reference-Format}
\bibliography{assets/bibs/papers}
\end{document}